\documentclass[11pt]{article}

\usepackage{graphicx,amsxtra}
\usepackage[usenames,dvipsnames,table]{xcolor}
\usepackage[letterpaper,margin=1in]{geometry}
\usepackage{verbatim}
\usepackage[authoryear]{natbib}
\usepackage{siunitx}
\usepackage{float}
\usepackage[hyphens]{url}

\newcommand{\ud}{\text{d}}

\let\originalleft\left
\let\originalright\right
\renewcommand{\left}{\mathopen{}\mathclose\bgroup\originalleft}
\renewcommand{\right}{\aftergroup\egroup\originalright}

\begin{document}

\title{The impact of species-neutral stage structure on macroecological patterns}


\author{Rafael D'Andrea$^{1, \ast}$ \& James P O'Dwyer$^{1, \dag}$ \\ 
  $^1${\small Dept.~Plant Biology, University of Illinois, Urbana-Champaign, IL, USA}\\  $^{\ast}${\small \texttt{rdandrea@illinois.edu}}
  $^{\dag}${\small \texttt{jodwyer@iillinois.edu}} \thanks{JOD acknowledges the Simons Foundation Grant \#376199, McDonnell Foundation Grant \#220020439, and Templeton World Charity Foundation Grant \#TWCF0079/AB47.}}

\maketitle

\begin{abstract}
\section*{}
Despite its radical assumption of ecological equivalence between species, neutral biodiversity theory can often provide good fits to species abundance distributions observed in nature. Major criticisms of neutral theory have focused on interspecific differences, which are in conflict with ecological equivalence. However, neutrality in nature is also broken by differences between conspecific individuals at different life stages, which in many communities may vastly exceed interspecific differences between individuals at similar stages. These within-species asymmetries have not been fully explored in species-neutral models, and it is not known whether demographic stage structure affects macroecological patterns in neutral theory. Here we present a two-stage neutral model where fecundity and mortality change as an individual transitions from one stage to the other. We explore several qualitatively different scenarios, and compare numerically obtained species abundance distributions to the predictions of unstructured neutral theory. We find that abundance distributions are generally robust to this kind of stage structure, but significant departures from unstructured predictions occur if adults have sufficiently low fecundity and mortality. In addition, we show that the cumulative number of births per species, which is distributed as a power law with a 3/2 exponent, is invariant even when the abundance distribution departs from unstructured model predictions. Our findings potentially explain power law-like abundance distributions in organisms with strong demographic structure, such as eusocial insects and humans, and partially rehabilitate species abundance distributions from past criticisms as to their inability to distinguish between biological mechanisms.

\vspace{1cm}
\textbf{Keywords}: Species abundance distribution, Demographic structure, Progeny distribution,  Neutral biodiversity theory
\end{abstract}

\newpage
\section{Introduction}
A long-standing goal of community ecology is to understand the connection between processes of community assembly and patterns of biodiversity. Neutral biodiversity theory (NBT) poses that such patterns are driven by ecological drift, i.e. stochastic birth and death events whose probabilities are irrespective of species identity \citep{Hubbell2001}. Despite evidence that deterministic factors such as habitat structure shape biodiversity \citep{Condit2011}, NBT successfully describes observed species abundance distributions (SADs), providing better fits than statistical distributions in tropical forests~\citep{Volkov2003,Harte2003}, and in some datasets even better than models incorporating niche differences~\citep{Etienne2005}. This renewed the debate over the relative role of niche partitioning versus stochastic neutral forces in shaping biodiversity patterns \citep{Ruokolainen2009,Vergnon2009,Chisholm2010,DAndrea2016b}. 

Species abundance distributions in tropical forests commonly approximate a logseries \citep{Bell2000,white2012characterizing}, which is well described by NBT \citep{Volkov2003} despite its radical assumption of demographic equivalence between individuals of different species. This postulated equivalence is manifestly broken between conspecific individuals at different life stages. In trees, seedling mortality is much higher than adult mortality, whereas fecundity increases considerably as an individual ages to larger size \citep{Harcombe1987}. For example, seed output in \emph{Cecropia obtusifolia} may differ by up to three orders of magnitude between juveniles and mature individuals, while mortality may change by a factor of 20 \citep{Alvarez-Buylla1992}. \citet{Rosindell2012b} have shown that NBT's fits for abundances of reproductive individuals are sensitive to non-random sampling bias caused by stage structure. However, the implications of stage structure for neutral theory predictions themselves have not been fully explored. \citet{ODwyer2009} showed that neutral SADs are robust to demographic structure in mortality. Here, in addition to mortality we explore the impact of ontogenetic differences in fecundity, which was not addressed in that study. 

We also explore a second macroecological pattern, commonly used in tests of neutrality in the social sciences: the progeny distribution. This is the cumulative number of births within a species over a sufficiently long period of time. Because the progeny distribution aggregates the state of the community across time, it could potentially provide a better test for alternative community dynamics processes than the SAD. In the social sciences the progeny distribution is amply studied empirically \citep{Hahn2003,Bentley2004}, as birth registries are more commonly available than the corresponding data in ecological systems. Recent theoretical work has shown that the progeny in an unstructured neutral birth-death process follows a power law distribution with a specific exponent of 3/2 \citep{ODwyer2017}. It is not known whether this result holds for structured populations.  

We present a simple two-stage structured neutral model, where individuals at different stages may differ in both fecundity and mortality rates. We numerically simulate our model and compare SADs and progeny distributions to predictions from unstructured neutral theory. We begin with conceptual scenarios where adult rates are either high or low relative to the corresponding juvenile rates. We examine the four qualitative possibilities: high/low adult mortality relative to juvenile mortality $\times$ high/low adult fecundity relative to juvenile fecundity. The classical unstructured neutral model, where adults and juveniles have identical rates, sits in the middle of these four. We then explore two additional scenarios describing organisms where stage structure is hypothesized to be particularly consequential for abundance distributions: humans and eusocial insects. For the latter group we compare simulation results with observations in nature. Finally, we test the robustness of our results to relaxing the assumption of neutrality.

\section{Methods}

Neutral theory assumes that multiple species compete symmetrically for a single resource. One common formulation of neutrality, known as the non-zero sum approach, approximates these non-linear competitive interactions by allowing a focal species to interact with the average abundances of all other species, rather than the specific set of species abundances \citep{Volkov2003,Etienne2007,ODwyer2014}. This results in a simplified mean field scenario where in each step the chosen species may gain (birth) or lose (death) one individual as in a random walk with fixed step size. Deaths are slightly more likely than births, but total community size is maintained by a constant inflow of new species from speciation. We use this approximation in our model.

Consider an unstructured neutral model for a closed metacommunity of steady-state size $J$, with stochastic birth, death, and speciation in the form of an influx of new species with rate $\theta=\nu J$, where each new species starts with abundance 1. The mean field master equation describing the probability that a species has abundance $N$ at time $t$, is \citep{ODwyer2014}
\begin{equation}
\left.\frac{\partial P}{\partial t}(N,t)\right|_{N>1}=(1-\nu)\,b\,[(N-1)P(N-1|t)-NP(N|t)]+d\,[(N+1)P(N+1|t)-NP(N|t)]\label{Punst}
\end{equation}
\noindent where $b$ and $d$ are fixed per capita birth and death rates, with $b=d$. The first term describes births that do not result in speciation, hence the factor $1-\nu$ (the equation describes dynamics after introduction through speciation). This equation can be solved to give the logseries species abundance distribution typical of unstructured neutral theory:
\begin{equation}
\langle S(N)\rangle\simeq\frac{\theta}{N}\left( 1-\frac{\theta}{J}\right)^N\label{SADEQN}
\end{equation}
\citet{ODwyer2017} showed that at late times, the expected number of species with $k$ birth events since their appearance in the community is 
\begin{equation}
\langle S(k)\rangle=\frac{2d}{\nu}\left(\frac{4db(1-\nu)}{\nu^2}\right)^{k-1}(-1)^{k-1}\binom{\frac{1}{2}}{k}\label{ProgenyEQN}
\end{equation}
In the social sciences this is known as the progeny distribution. For intermediate values of $k$, the expression above behaves like a power law with a specific exponent of 3/2. For higher values exceeding $(b/\nu)^2$, it decays exponentially \citep{ODwyer2017}.

Our two-stage neutral model is analogous to the one described by Equation \ref{Punst}, except that in addition to births and deaths there are also aging events, when individuals in stage 1 transition to stage 2. We were unable to solve the master equation for the stage-structured model (provided in the supplementary information) given general birth and death rates of the two classes, so we take a simulation-based approach instead. Denoting by $W^+_{i}$ the rate at which a species subpopulation at stage $i$ with $N_i$ individuals gains an individual, and $W^-_{i}$ the rate at which it loses an individual, we write 
\begin{align}
W^+_1 &= (1-\nu)(b_1N_1+b_2N_2) & W^-_1 &= (a+d_1)N_1\label{W1}\\
W^+_2 &= aN_1 & W^-_2 &= d_2N_2\label{W2}
\end{align}
where $b_i$ and $d_i$ are respectively the fecundity and mortality of individuals at stage $i$, identical across species, and $a$ is the rate at which individuals from stage 1 transition to stage 2. In addition, speciation occurs at fixed rate $\nu(b_1\mathcal{N}_1+b_2\mathcal{N}_2)$, where $\mathcal{N}_i$ are the equilibrium abundances of stages 1 and 2 summed across species. We will refer to stages 1 and 2 respectively as juveniles and adults, without necessarily implying the actual biological connotations of those terms (our ``juveniles'' are allowed to reproduce, for instance). Figure \ref{scheme} shows a diagram of our model.

This model has a stationary state if $d_2>0$ and $(1-\nu)(b_1+b_2/d_2)<1+d_1$. In that equilibrium, the time-averaged total adult population will be proportional to the time-averaged total juvenile population, $\bar{N}_1=d_2\bar{N}_2$. Furthermore, the time-averaged population of stage $i$ will be $\mathcal{N}_i$ if $\mathcal{N}_1=d_2\mathcal{N}_2$ and $(b_1+b_2/d_2)=1+d_1$. We therefore choose as our free parameters juvenile fecundity ($b_1$), juvenile mortality ($d_1$), and adult mortality ($d_2$), and set adult fecundity to $b_2=(1+d_1-b_1)\,d_2$. With this parametrization, near the equilibrium a fraction $\nu$ of recruitment events are speciation events as opposed to local births, and the total community size is $\mathcal{N}_1+\mathcal{N}_2=(1+d_2)\,\mathcal{N}_2$. To ensure comparable community size across simulations, we set $\mathcal{N}_2=J/(1+d_2)$, with $J=110,000$ individuals. We set $\nu=0.001$ throughout our simulations. These parameter choices ensure high species diversity and community size, thus justifying our mean field approximation. (A community with endogenous speciation, of which our model is an approximation, would eventually go extinct. However, the timescale for decay to the absorbing state will be enormous for a large community, thereby justifying our approximation.) Notice that our fecundity and mortality parameters are defined relative to the aging rate between the juvenile and adult stage, which we arbitrarily set to 1 by rescaling time to the appropriate unit. In other words, for every juvenile that ages into adulthood, $d_1$ juveniles die, etc. This choice does not affect abundance distributions.

\subsection*{Conceptual scenarios}
\begin{figure}[H]
\caption{Placement of our various simulation scenarios in parameter space, defined by adult mortality and fecundity relative to juvenile mortality and fecundity. Classical unstructured neutral theory falls in the middle. Arrows indicate that actual placement falls outside the bounds of the diagram.Scenario abbreviations are JT: juvenile turnover, AT: adult turnover, JV: juvenile vigor, AV: adult vigor, CM: Children of Men, EI: eusocial insect model, UNST: unstructured.} \label{Cartoon}
\includegraphics[width=.5\textwidth,angle=0]{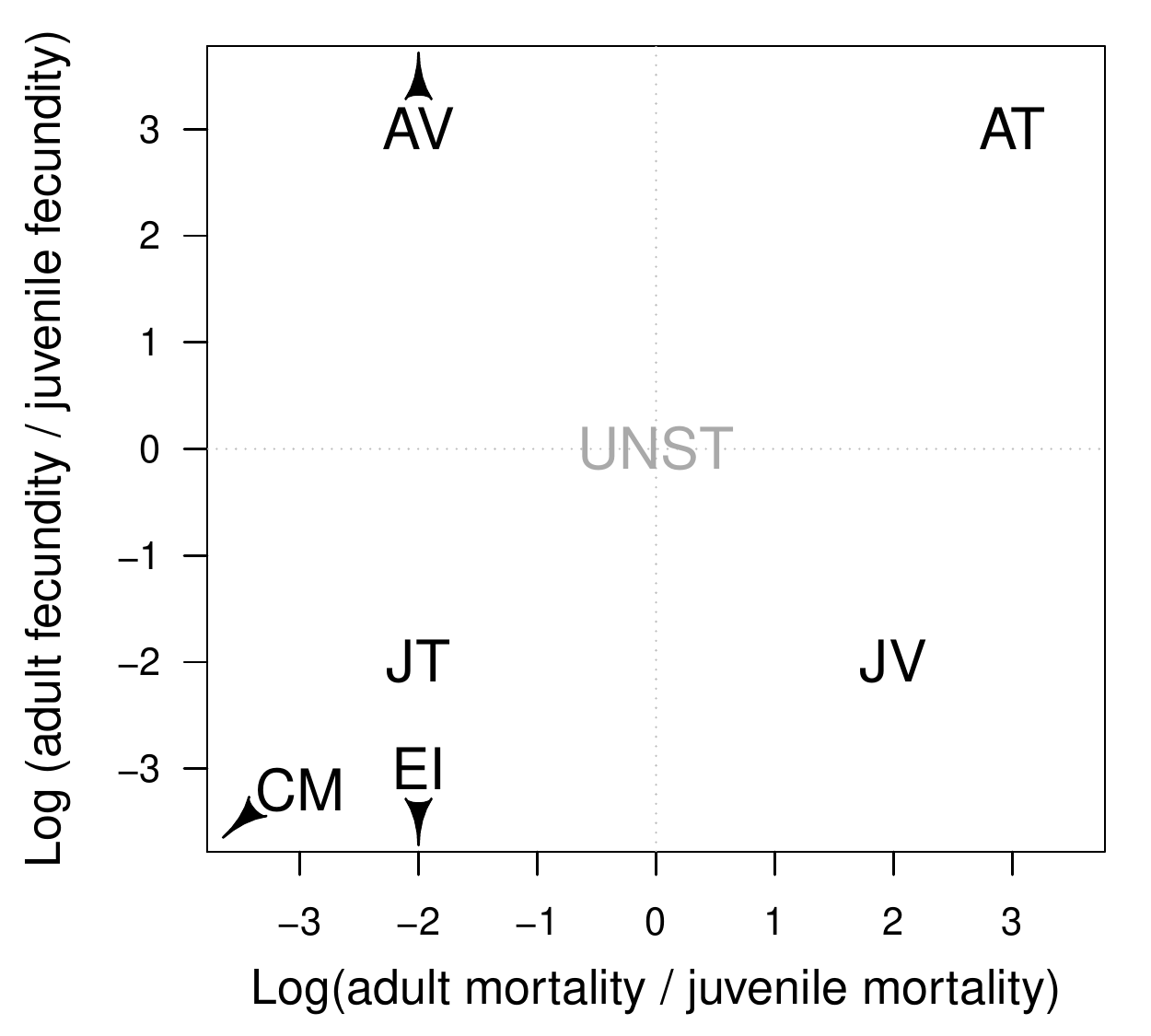}
\centering
\end{figure}

In the Juvenile Turnover (JT) scenario, adults have low mortality and fecundity relative to juveniles, $d_2/d_1=10^{-2},\;\; b_2/b_1=10^{-2}$. The Adult Turnover (AT) scenario is the reverse: $d_2/d_1=10^3,\;\; b_2/b_1=10^3$. In the Juvenile Vigor scenario (JV), adults have high mortality and low fecundity relative to juveniles: $d_2/d_1=10^2,\;\; b_2/b_1=10^{-2}$. Finally, the Adult Vigor (AV) scenario is the opposite: $d_2/d_1=10^{-2},\;\; b_2/b_1=\infty$ (juveniles do not reproduce). Many organisms will fall somewhere between these four scenarios. For example in forest trees, high-mortality pre-reproductive saplings are followed by low-mortality reproductive adults, similar to scenario AV. Moths and butterflies with well-defended caterpillars pupating into adults under high predation would be closer to AT, as are rapidly senescing organisms such as Pacific salmon and annual plants. Placental mammals where senescence is more gradual or which experience menopause, such as great apes and cetaceans, could be better described by JT or JV. The unstructured scenario, where adults and juveniles have identical rates, is in the middle of these four (Figure \ref{Cartoon}; see Table \ref{table1} for our parameter choices in each scenario).

\subsection*{Children of Men scenarios}
In humans, most reproduction occurs early in life, followed by a long-lasting non-reproductive stage. Human given names follow a 3/2-power law distribution rather than a logseries, similar to the distribution of baby names \citep{Bentley2004}. In our two-stage model, the master equation becomes solvable in the limit where adults are sterile and immortal, and we find that the distribution of adults approximates a power law with exponent 3/2, similar to the behavior of the progeny in the unstructured model (supplementary information). Based on this result, we hypothesize that when adults have low mortality and fecundity relative to juveniles, the abundance distribution will depart from the logseries and move towards a 3/2-power law. We test this hypothesis on a set of simulation scenarios where we set adult fecundity very low relative to juveniles, and turn adult mortality to increasingly low levels (see Table \ref{table1}). We call these the Children of Men scenarios by analogy with P. D. James's (\citeyear{James1992}) dystopian novel where people can no longer have children. 

\subsection*{Eusocial insect scenarios}
\begin{figure}[H]
\caption{Stage-structured model for eusocial insects. Populations are composed of reproductive queens and sterile workers. Queens undergo a stochastic birth-death process with sporadic speciation. Worker subpopulation receives input from the queen subpopulation, but does not feed back into it. Notice that queens do not ``age'' into workers.} \label{FluxANT}
\includegraphics[width=.5\textwidth,angle=0]{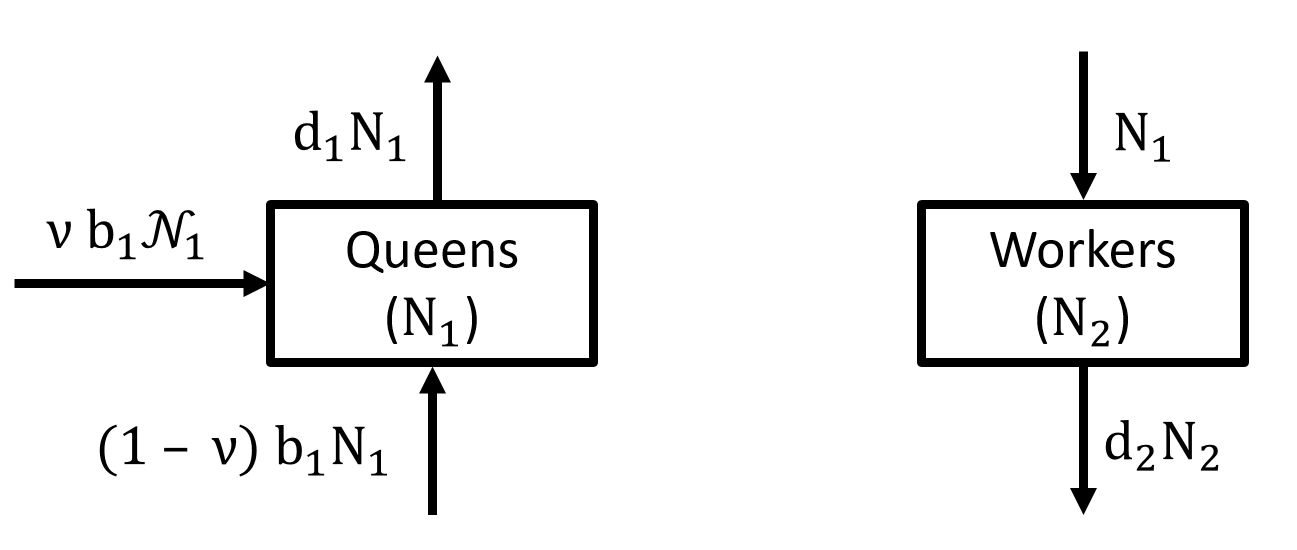}
\centering
\end{figure}

Eusocial insects present one of the most striking examples of demographic structure in nature, characterized by a large number of sterile workers and one or few reproductive queens. Given the resemblance of this demographic structure to our Children of Men scenarios, we adapted our two-stage model to a simplified representation of eusocial insects (Figure \ref{FluxANT}). Our eusocial insect model consists of a reproducing queen caste and a non-reproducing worker caste. The queen population undergoes a neutral birth-death process with speciation, whereas the worker subpopulation receives input from but does not feed back into the queen subpopulation, i.e. $b_2=0$. Queens produce workers at rate 1, which can always be done by scaling time appropriately. We set $d_1=b_1$, and therefore in stationary equilibrium $N^*_1=\mathcal{N}_1=d_2 N^*_2$. In addition to complete sterility in one of the stages, a major difference from our model above is that queens do not ``age'' into workers. This is therefore not a case of stage structure in the sense of ontogenetic stages, but demographic structure in the sense that there are different classes of individuals in the same species.

\subsection*{Distribution fitting}
In order to compare results with predictions from the unstructured neutral model, we used maximum likelihood to fit a logseries to the SAD (Equation \ref{SADEQN}, see also \citealt{Alonso2008}) and a truncated power law to the progeny, holding above a lower bound \citep{Gillespie2015}. The latter is a simplified approximation to the analytical prediction (Equation \ref{ProgenyEQN}). The fitted power law exponent is sensitive to the lower bound, and to estimate it we use the method of minimum Kolmogorov-Smirnov distance \citep{Clauset2009}. We coded all simulations in R \citep{RcoreTeam2017}, and fitted power laws and logseries using R packages ``poweRlaw'' and ``sads'' \citep{Clauset2009, sads2016}. R code for fitting power laws to large numbers is provided in https://github.com/odwyer-lab/Power-laws-in-large-data.

\section{Results}

\begin{figure}[H]
\caption{\textbf{Top} and \textbf{middle} rows: Cumulative species abundance distribution and progeny distribution in each of our four two-stage scenarios (black dots), overlaid with maximum likelihood fits for a logseries (SAD, blue curves) and power law (progeny, red lines). Chi-squared tests on SADs indicate a very good fit to a logseries in each scenario (p-value $= 1$). Maximum likelihood estimates for the power law exponent, $\hat{\alpha}$, approximate 3/2 in all scenarios. \textbf{Bottom}: Juvenile and adult populations are roughly proportional to each other as $N_{1a}\approx d_2N_{2a}$, as seen by plotting $\textrm{log}(1+N_{1a})$ against $\textrm{log}(1+d_2N_{2a})$ for each species at the end of the simulation. Linear regression fits (green line) are very close to the one-to-one line (in red). Legend shows adjusted R-squared of the linear regression.} \label{sadspgns}
\includegraphics[width=1\textwidth,angle=0]{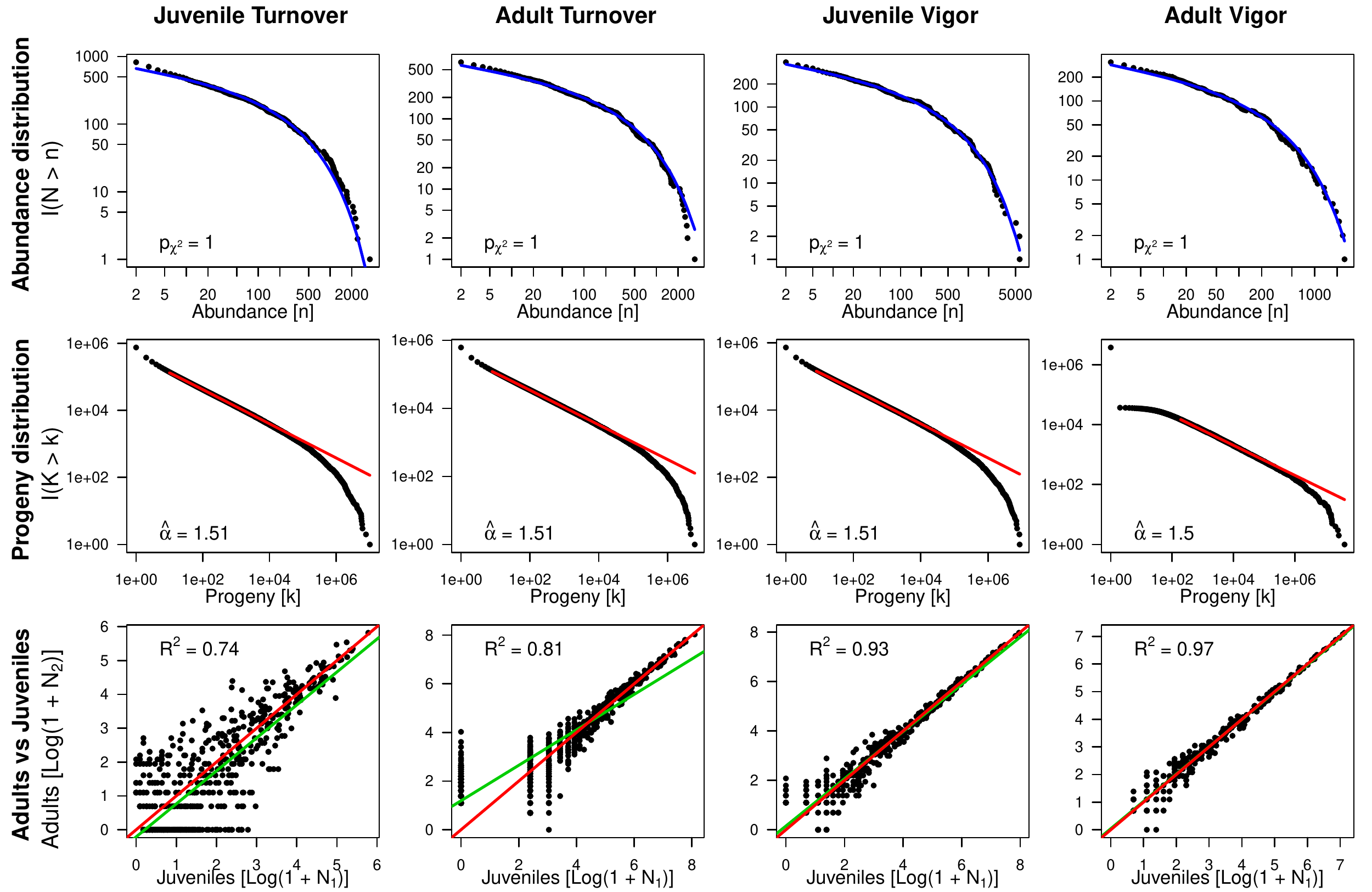}
\centering
\end{figure}

The top and middle rows of Figure \ref{sadspgns} show the cumulative species abundance distribution and progeny distribution, respectively, in the JT, AT, JV, and AV scenarios. In all four, the SAD fits a logseries and the progeny fits a power law with exponent close to 3/2. This occurred even though between those scenarios adult mortality and fecundity differed from juveniles by a factor of up to 1,000 (see Table \ref{table1}). The bottom row of Figure \ref{sadspgns} shows the relationship between juvenile and adult populations across species, respectively. In all four scenarios those populations are proportional to each other, with $N_{1}\approx d_2N_{2}$. According to Equation \ref{W2}, this is precisely the equilibrium condition for the adult population. In fact, in all four scenarios the coefficient of variation of both the juvenile and adult populations through time are no bigger than that of the community as a whole, indicating that each life stage is in equilibrium separately and synchronously. Supplementary Figure S1 shows community size, richness, and the synchrony between the populations of the two life stages. 

\begin{figure}[H]
\caption{Children of Men scenarios. \textbf{Top}: As adult mortality decreases, the SAD gradually deviates from a logseries towards a power law. Goodness-of-fit $\chi^2$ p-values quickly drop to significant values, and the logseries fit (blue curve) is thus rejected. In contrast, the exponent $\hat{\alpha}$ of the fitted power law (red curve) approaches 3/2. \textbf{Middle}: Unlike the SAD, the progeny distribution is invariant across these scenarios, remaining a good fit to a 3/2 power law. \textbf{Bottom}: As adult mortality is lowered, the proportionality between juvenile and adult populations deteriorates. Axes, lines, and legends as in Figure \ref{sadspgns}.} \label{CoM}
\includegraphics[width=1\textwidth,angle=0]{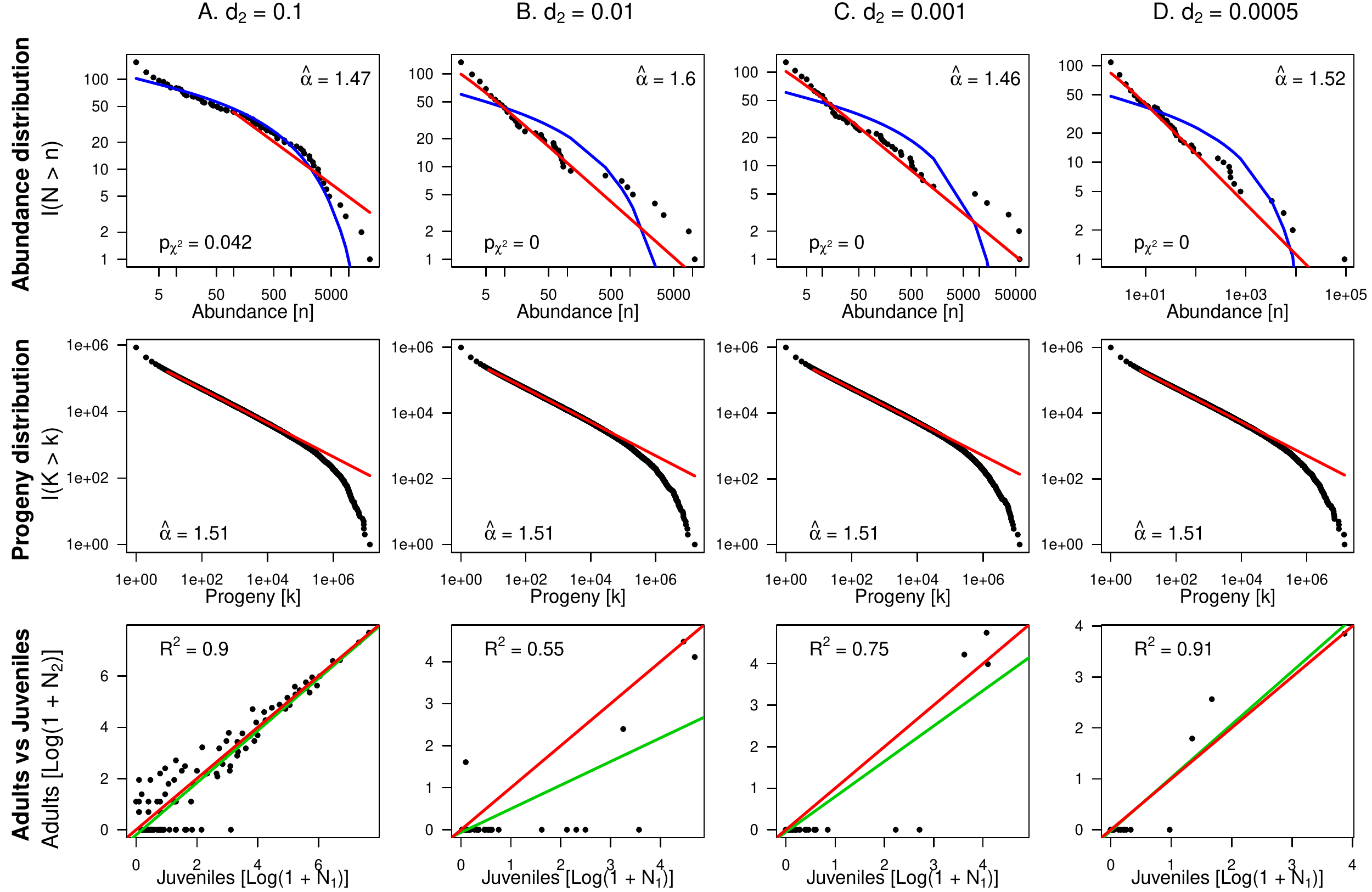}
\centering
\end{figure}

In the Children of Men scenarios, as we dial down adult life history rates, SADs depart from a logseries distribution (Figure \ref{CoM}, top row). The progeny distribution, on the other hand, remains a good fit to a 3/2 power law (Figure \ref{CoM}, middle row). When adult mortality is very low, the SAD fits a 3/2 power law better than a logseries (Figure \ref{CoM}, top row), confirming expectations from our analytical result (supplementary information). We explain this as follows. When adults do not die, the number of living adults approximates the cumulative number of births, i.e. the progeny. Also because of their low mortality, adults vastly outnumber juveniles, and hence dominate the SAD. The end result is to push the SAD towards a progeny-like distribution, namely a 3/2 power law. Note also that as adult mortality decreases, adult and juvenile populations break out of lockstep, and $N_1=d_2N_2$ is no longer a good fit (Figure \ref{CoM}, bottom row; Supplementary Figure S2 shows the loss of synchrony between the two life stages).

\begin{figure}[H]
\caption{Eusocial insect scenarios. Plot layout similar to Figure \ref{CoM}.} \label{ANT}
\includegraphics[width=.8\textwidth,angle=0]{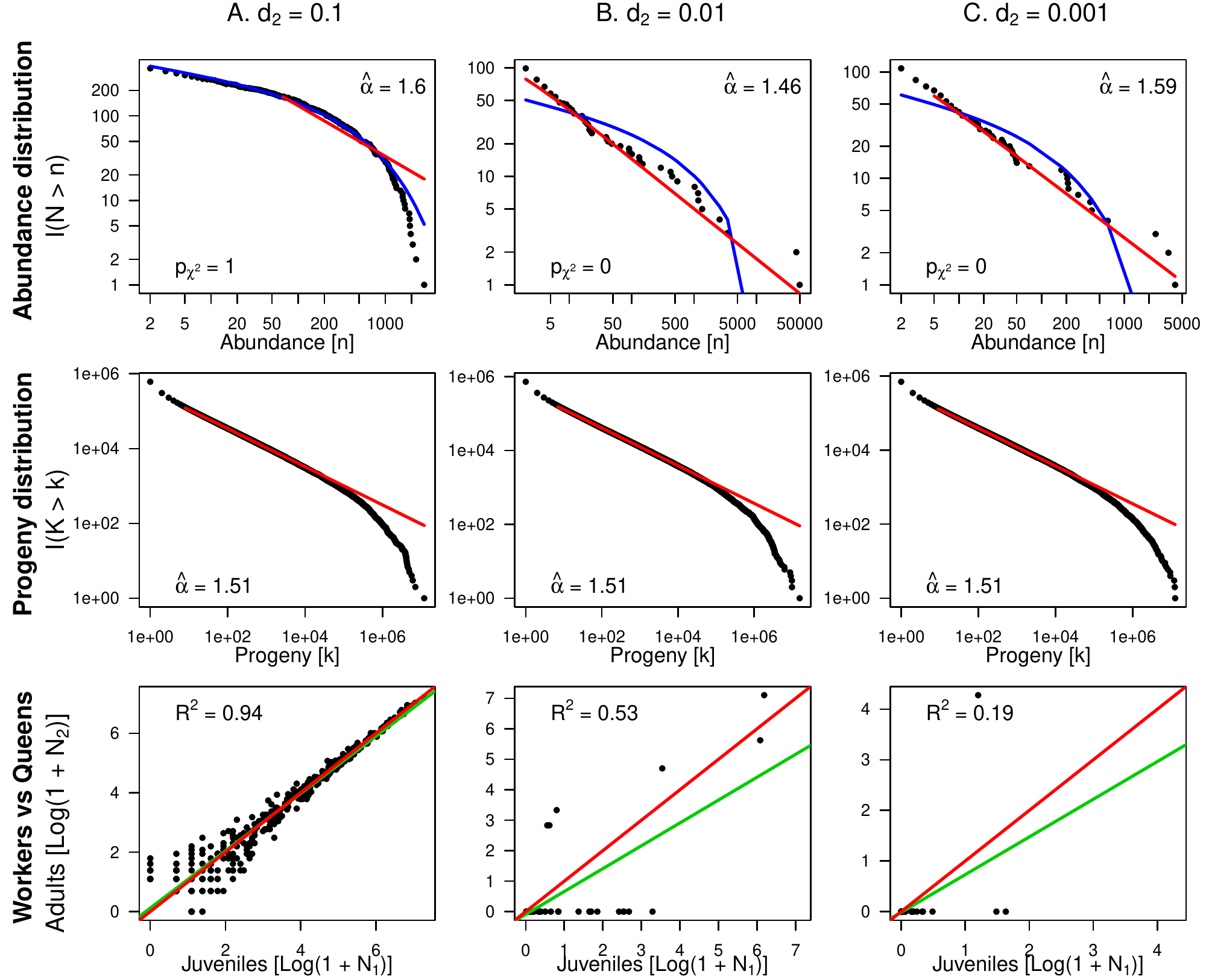}
\centering
\end{figure}

Our eusocial insect model displays similar behavior to our Children of Men scenario (Figure \ref{ANT}): when worker mortality is sufficiently high, the SAD fits a logseries, and queen and worker subpopulations are mutually proportional. But as worker mortality is reduced, the linear relationship breaks down (see also Supplementary Figure S3), and the SAD veers towards a progeny-like 3/2 power law. The progeny distribution, as in all other scenarios, is invariant throughout. 

\subsection*{Power law abundance distribution in insect groups}
\begin{figure}[H]
\caption{Abundance distribution of eight different arthropod groups. Blue and green curves are maximum likelihood fits to a logseries and power law with exponential cutoff, respectively. Legends indicate the fitted exponent of the power law phase, and the p-value of the likelihood ratio test between the two fits, with $p< 0.05$ indicating that the power law with cutoff is significantly better than the logseries. $S$ is the number of species sampled in each group, and $J$ the number of individuals.} \label{InsectPlots}
\includegraphics[width=1\textwidth,angle=0]{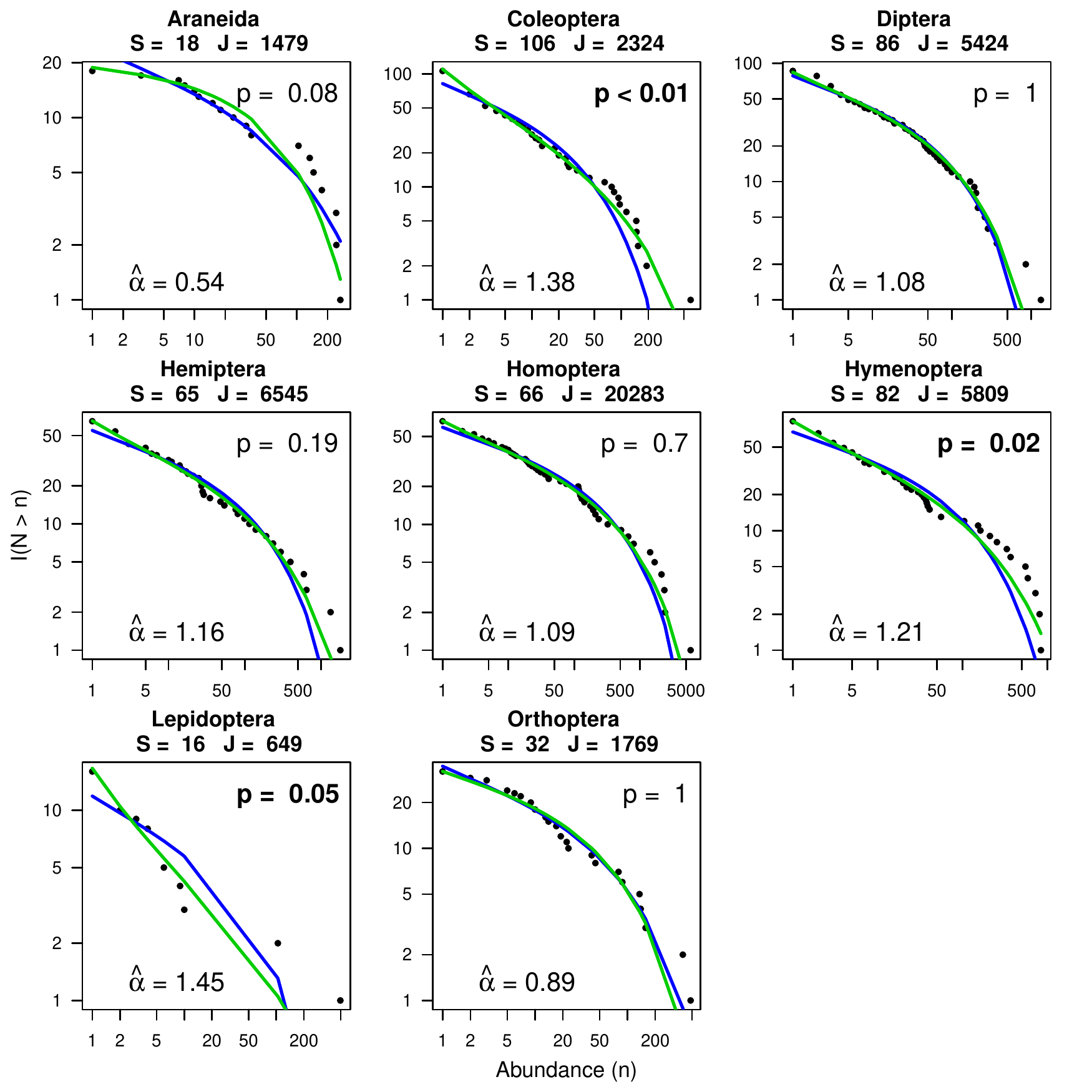}
\centering
\end{figure}

\citet{Siemann1999} found a power law relationship between species rank and abundances across several arthropod groups, with an exponent often approximating 2. This suggests a power law SAD with exponent near 3/2 (see supplementary information). Using data available from the University of Minnesota Cedar Creek data repository \citep{Haarstad2004}, we fitted a logseries and a power law with exponential cutoff to abundance distributions of eight arthropod taxa, including Hymenoptera, a diverse order of insects containing many eusocial species (Figure \ref{InsectPlots}). We fitted using maximum likelihood estimation and estimated significance with a likelihood ratio test. Given the multiple tests we adjusted p-values using Bonferroni correction. We found that the power law with exponential cutoff is a significantly better fit than the logseries in Coleoptera, Hymenoptera, and marginally Lepidoptera, although the data contained very few species in the last group, inviting caution in interpreting results. (In the case of Coleoptera, even a simple power law fits better than the logseries, and the fitted exponent is a suggestive $\hat{\alpha}=1.52$.) Although these results are by no means conclusive, the fact that orders containing eusocial insects showed the strongest deviation from a logseries SAD suggests the possibility that the stage-structured neutral model captures some of the essential drivers of abundance distributions in taxa with this demographic structure. 

\subsection*{Relaxing neutrality}
We tested the sensitivity of our results to the neutrality assumption using a scenario where species are so different from one another that each occupies their own niche---thus the opposite of neutrality, where all species occupy the same niche (see details of the model supplementary information). Similar to the neutral model, adding stage structure in this model typically did not change either the SAD or the progeny. However, unlike the neutral model, we did not observe changes to the SAD even in Children-of-Men-like scenarios when adults had very low mortality and fecundity relative to juveniles (see Supplementary Figure S4).

\section{Discussion}
Neutral theory often succeeds in describing macroecological patterns such as species abundance distributions. Although criticisms and expansions of the theory generally focus on interspecific differences, neutrality can also be broken by dramatic differences between conspecific individuals at different life stages. Here we investigated how such differences impact predictions on the distribution of abundances (SAD) and cumulative births (progeny). Our models span a variety of scenarios with two life stages, where these stages are distinguished by their fecundity and/or mortality rates. Our results indicate that both abundance and progeny distributions are largely insensitive to demographic variation in either fecundity or mortality, as long as as differences in individuals' fecundity and mortality pertain strictly to their life stage and not their species identity. However, SADs can deviate from the unstructured neutral community's logseries if both fecundity and mortality drop significantly at later life stages. The progeny distribution, in contrast, seems truly invariant under a wide variety of species-neutral demographic structure. \citet{ODwyer2017} showed that in the stationary equilibrium of a neutral birth-death process with speciation, the progeny follows a power law with exponent 3/2, followed by an exponential drop-off at very large values. This was also observed in all scenarios of our stage-structured neutral model.

\begin{figure}[ht!]
\caption[Scheme]{Schematic representation of the 2-stage model, showing the flow of individuals into and out of each stage. In the stationary state, $N_1=N_1^0$, $N_2=N_2^0=1/d_2N_1^0$, and a proportion $1-\nu$ of birth events are local births, and the remainder are speciation events. For simplicity, we set the time scale so that the aging rate is normalized to 1. Parameters $b_1$ (juvenile fecundity), $d_1$ (juvenile mortality), and $d_2$ (adult mortality) are independently varied, while adult fecundity is set to $b_2=(1+d_1-b_1)\,d_2$. A stationary state exists only if $d_2>0$ and $b_1<1+d_1$. } \label{scheme}
\includegraphics[width=.5\textwidth,angle=0]{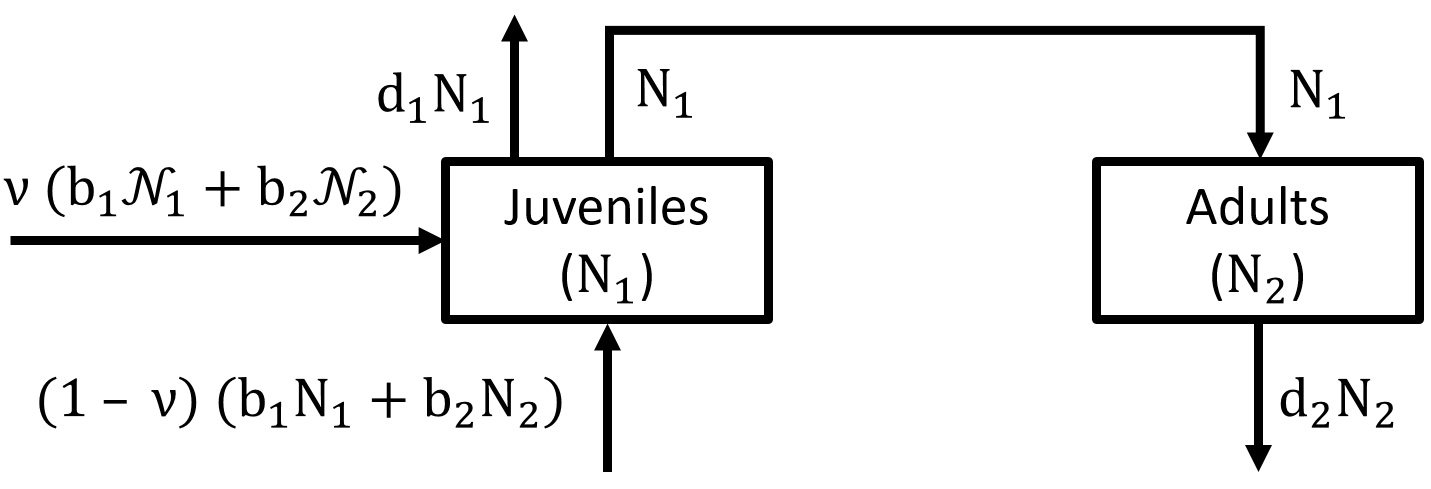}
\centering
\end{figure}

Our results for how and when the neutral SAD departs from the log series distribution have intuitive explanations. First, in scenarios where adult and juvenile populations are typically proportional to each other, our models effectively reduce to a simpler, unstructured birth-death process. A glance at Figure \ref{scheme} reveals that if $N_{1a}(t)=k N_{2a}(t)$ with constant $k$, then adults undergo a neutral birth-death process with death rate $d_2$ and birth rate $k$, while juveniles undergo a neutral birth-death process with rates $1+d_1$ and $(1-\nu)(b_1+b_2/k)$, and therefore so does the species population as a whole. This fact explains why the SAD is largely unchanged in our first four scenarios. 

However, if adult mortality is too low, the equilibrium juvenile population is so small that stochastic fluctuations become important, and break the synchrony between the sub-populations in either stage. Even though each subpopulation may still be undergoing a simple birth-death process, the species as a whole is no longer doing so, and as result the SAD is affected. The greater robustness of the progeny distribution is likely due to its independence of death rates. The SAD on the other hand depends on the balance between births and deaths, which our results show can be broken if intraspecific subpopulations with different life history rates do not fluctuate in synchrony. Our Children of Men scenario provides an example of this, and may provide an explanation of the kinds of progeny and abundance distributions that previous research has identified in the social sciences. For example, both baby names (the progeny distribution) and censuses of names (analogous to the SAD) approximately follow a power law. Our study hints that this may reflect a particular stage-structured neutral dynamics in human names, possibly caused by the fact that in human societies most childbearing is done by young adults, while those no longer reproducing still have many years to live, and hence appear in census counts.

This Children of Men scenario may also have a biological counterpart in groups of animals where the population contains many non-reproducing adults, such as eusocial insects. We presented a modified version of our two-stage neutral model adapted for eusocial insects, and showed that the SAD approximates a 3/2 power law for sufficiently low worker mortality. \citet{Siemann1999} found that species rank-abundance distributions across arthropod groups approximate a power law with exponent close to $2$, which is equivalent to a $3/2$ power-law abundance distribution (see supplementary information). Using a similar dataset collected by the same authors, we found that out of eight different arthropod taxa, a power law with exponential cutoff is a significantly better fit than a logseries for the SADs of Coleoptera and Hymenoptera. Concerning the latter group, we speculate that this may be related to the common occurrence of eusociality in Hymenoptera. It is unlikely that eusocial insects undergo neutral dynamics. But the observation that abundance distributions approximate predictions from a stage-structured neutral model implies that neutrality (even if inaccurate) may be an effective vehicle for capturing the effects of stage structure on SADs. A similar empirical study with larger numbers of individuals and larger numbers of species would be needed to settle this conjecture. A count of both colonies and individuals (as in \citealt{Ernest2009}) is desirable, as a definitive result would be if the colony numbers undergo standard drift (logseries), while the SAD approximates a 3/2 power law.

Our finding that SADs are generally insensitive to stage structure whether species are under complete niche overlap (neutrality) or zero niche overlap (when each species occupies their own niche) suggests that this robustness will also occur in scenarios with intermediate niche overlap. On the other hand, strong stage structure of the Children-of-Men type affects the SAD in the neutral model but not in our zero-overlap scenario. Of course, the landscape of niche-overlap scenarios is huge, and between complete overlap and no overlap there are intermediate niche theories which do display the neutral type results (see e.g. \citealt{Chisholm2010}). We conclude that at some point between complete neutrality and a niche for each species, niche structure becomes the predominant driver of species abundances, even under very strong stage structure. 

Our simple two-stage model is of course a very simplified representation of actual demographic structure in nature, where individuals undergo more than two distinct life stages differing substantially in life history. However, we note that our argument for the robustness of the SAD holds regardless of the number of stages, so long as sub-populations vary in relative synchrony. In our model, competitive interactions between individuals are identical irrespective of their life stages. In reality, competition may be stronger at some stages than at others. For example among tropical trees most of the competition for light is between saplings in the understory.

It has been argued that abundance distributions are consistent across different systems because they are not sensitive to any particular community's underlying ecology, thus being a poor test of community assembly processes \citep{McGill2003}. Our study qualifies this, suggesting that abundance distributions are affected by extreme stage structure, as in eusocial insects and species with long-lived non-reproducing adults such as humans. When stage structure is not as marked, species abundance distributions may indeed be insensitive to it, and thus uninformative of it. The success of neutral theory in reproducing observed abundance distributions may imply that it is an effective or sufficient model of the ultimate drivers of abundance patterns, despite omitting biological complexities which are critical for understanding the success or failure of individual species.

This is the first study to systematically explore intraspecific variation in fecundity, and demographic structure more generally, in the context of neutrality. The fact that it can strongly impact one of the most widely tested patterns under neutral theory is noteworthy. 

\vspace{1cm}
\textbf{Acknowledgments}:
The authors thank Evan Siemann and David Tilman for generously agreeing to our use of their data in this paper. The data are available at \url{http://www.lter.umn.edu/research/data/experiment?e122}. JOD acknowledges the Simons Foundation Grant \#376199, McDonnell Foundation Grant \#220020439, and Templeton World Charity Foundation Grant \#TWCF0079/AB47.


\newpage
\section*{Tables and Figures}

\begin{table}[!h]
\caption{Parameter values used in each of first set of two-stage scenarios. Mortality and fecundity of stage $i$ are represented respectively by $d_i$ and $b_i$, while $\mathcal{N}_i$ is the average total population size at stage $i$ (given in thousands). Speciation rate $\nu=0.001$ is fixed across scenarios. Scenario abbreviations are HJT: high juvenile turnover, HAT: high adult turnover, ASE: adult senescence, AVI: adult vigor, COM: Children of Men, ANT: eusocial insect model.}
\label{table1}     
\resizebox{\textwidth}{!}{\begin{tabular}{llllllllllll}
\hline\noalign{\smallskip}
\textbf{} & \textbf{HJT} & \textbf{HAT} & \textbf{ASE} & \textbf{AVI} & \textbf{COMa} & \textbf{COMb} & \textbf{COMc} & \textbf{COMd} & \textbf{ANTa} & \textbf{ANTb} & \textbf{ANTc}\\
\noalign{\smallskip}\hline\noalign{\smallskip}
\textbf{$d_1$} & 10 & 0.01 & 0.01 & 100 & 0.01 & 0.01 & 0.01 & 0.01 & 1 & 1 & 1\\
\textbf{$b_1$} & 10 & 0.01 & 1 & 0 & 1 & 1 & 1 & 1 & 1 & 1 & 1\\
\textbf{$d_2$} & 0.1 & 10 & 1 & 1 & 0.1 & 0.01 & \num{1e-3} & \num{5e-4} & 1 & 0.01 & \num{1e-3}\\
\textbf{$b_2$} & 0.1 & 10 & 0.01 & 101 & \num{1e-3} & \num{1e-4} & \num{1e-5} & \num{5e-6} & 0 & 0 & 0\\
\textbf{$\mathcal{N}_1$} & 10 & 100 & 55 & 55 & 10 & 1 & 0.1 & 0.05 & 55 & 1 & 0.1\\
\textbf{$\mathcal{N}_2$} & 100 & 10 & 55 & 55 & 100 & 109 & 109.9 & 109.95 & 55 & 109 & 109.9\\
\noalign{\smallskip}\hline
\end{tabular}}
\end{table}

\newpage

\newpage

\newpage

\newpage

\newpage

\section*{Supplementary Information}
\appendix
\renewcommand\thefigure{S\arabic{figure}}    
\setcounter{figure}{0}    

\section*{Appendix 1: Stage-structured neutral model}
The master equation for our stage-structured model is 
\begin{eqnarray}
\frac{\partial P}{\partial t}(N_1,N_2,t)&=&
(N_1+1)\,P(N_1+1,N_2-1,t)-N_1\,P(N_1,N_2,t)\nonumber\\
&+& (1-\nu)[b_1(N_1-1)\,P(N_1-1,N_2,t)-b_1N_1\,P(N_1,N_2,t)]\nonumber\\
&+& (1-\nu)[b_2N_2\,P(N_1-1,N_2,t)-b_2N_2\,P(N_1,N_2,t)]\nonumber\\
&+& d_1(N_1+1)\,P(N_1+1,N_2,t)-d_1N_1\,P(N_1,N_2,t)\nonumber\\
&+& d_2(N_2+1)\,P(N_1,N_2+1,t)-d_2N_2\,P(N_1,N_2,t)\label{ME}
\end{eqnarray}
where the first line corresponds to aging events with rate arbitrarily set to 1, the second and third lines to non-speciation birth events originating from juveniles and adults respectively, and the last two lines correspond to deaths of juveniles and adults, respectively.

The generating function for $P(N_1,N_2,t)$ is defined as $G(x,y,t) = \sum_{N_1,N_2} P(N_1,N_2,t)\, x^{N_1} y^{N_2}$. Substituting Equation \ref{ME}, we obtain:
\begin{equation}
\frac{\partial G}{\partial t} = \left[(1-\nu)b_1x(x-1)-d_1(x-1) - (x-y)\right] \frac{\partial G}{\partial x}+\left[(1-\nu)b_2y(x-1)-d_2(y-1)\right]\frac{\partial G}{\partial y}\label{G}
\end{equation}
The initial condition is a speciation event and consists of one juvenile and zero adults, and therefore $G(x,y,0)=x$. From the normalization of the probability, we get the boundary condition $G(1,1,T)=1$.

We were unable to find a solution to this equation. However, the Children of Men scenario where adults are an inert class ($b_2=d_2=0$), is more tractable. Because adults do not feed back into juveniles, the latter essentially undergo a fixed-rate birth-death process. And as adults do not die, the adult count will approximate the cumulative number of births in the species, i.e. the progeny (except for the juveniles that die without aging, which will be a small fraction if juvenile mortality is low compared to aging rate). We therefore expect the distribution of adults to mirror the progeny distribution of a neutral birth-death process, which has been recently shown to follow a power law with exponent 3/2 \citep{ODwyer2017b}. As $b_2$ and $d_2\rightarrow 0$, the equilibrium condition for juveniles converges to $d_1=(1-\nu)b_1-1$. Defining $b=(1-\nu)b_1$ and substituting in Equation \ref{G}, we can solve it using the method of characteristics \citep{Kendall1948}. The general solution is 
\begin{equation}
G(x,y,t)=\frac{\alpha(\beta-x)+\beta(x-\alpha)e^{-b(\beta-\alpha)t}}{(\beta-x)+(x-\alpha)e^{-b(\beta-\alpha)t}}
\end{equation}
where $\alpha$ and $\beta$ are the $x$-roots of 
\begin{equation}
bx(x-1)-(b-1)(x-1) - (x-y)=0
\end{equation}
with $\beta>\alpha$.

At large times $t\rightarrow \infty$ we get 
\begin{equation}
G(x,y,\infty)=\alpha=1-\sqrt{\frac{1-y}{b}}
\end{equation}
Notice that the expression is independent of $x$. This indicates that eventually all juveniles die and only adults remain. The Taylor coefficients of $G$ around $y=0$ provide the probability distribution of adults at large times:
\begin{equation}
P(N_2=n,\infty)=\frac{1}{\sqrt{b}}(-1)^{n-1}\binom{1/2}{n}
\end{equation}
for $n>0$, and $P(0,\infty)=1-1/\sqrt{b}$. This expression is similar to the progeny distribution derived in \citep{ODwyer2017b}, confirming our expectation of a relation between these distributions. Specifically it contains the same binomial coefficient, which imparts the power-law-like behavior for $n\gg 1$. The exponent of the power law, 3/2, does not depend on any model parameters.

\newpage
\section*{Appendix 2: Relaxing neutrality}

Here we present a stage-structured model where neutrality between species is relaxed. We do so by introducing a non-linearity in juvenile mortality. Here the rate at which the juvenile subpopulation loses an individual is 
\begin{equation}
W_1^-=(a+d_1N_1)N_1
\end{equation}
(compare with Equation 4 in the main text). As before, we set the per capita aging rate $a=1$. All other rates, $W_1^+$, $W_2^+$, $W_2^-$ are the same as in our neutral model. The nonlinearity in juvenile death introduces self-regulation: as the juvenile population grows, intraspecific competition leads to an increase in per capita mortality. Notice that species only regulate themselves and no others. This corresponds to a scenario where species are so different in their needs and strategies that there is no interspecific competition. In effect, each species occupies their own niche. 

The introduction of stabilization leads to a carrying capacity $N_1^*$, as the juvenile subpopulation will have a negative balance when $N_1>N_1^*$ and a positive balance when $N_1<N_1^*$. Solving the model for equilibrium gives $N_1^*=(b_1+b_2/d_2-1)/d_1$ and $N_2^*=N_1^*/d_2$. (Here we set $\nu=0$, eliminating the possibility of speciation, otherwise the community grows without bounds). 

We consider three scenarios: Equal Fecundity (EF), where juveniles and adults have the same fecundity $b_1=b_2$; Sterile Juveniles (SJ), where $b_1=0$; and Sterile Adults (SA), where $b_2=0$. The other free parameters in the model are set by choosing the carrying capacity $Q=N_1^*+N_2^*$, the proportion of juveniles in the species equilibrium $\rho=N_1^*/Q$, and the ratio between per capita juvenile mortality and per capita adult mortality $\varepsilon=d_1N_1^* /d_2$. We set $\rho_{EF}=0.5$, $\varepsilon_{EF}=1$, $\rho_{SJ}=0.8$, $\varepsilon_{SJ}=10^3$, $\rho_{SA}=0.1$, $\varepsilon_{SA}=10^3$. With these parameter choices, scenario SJ is qualitatively analogous to the Adult Vigor scenario in our neutral model (adults have high fecundity and low mortality relative to juveniles), and scenario SA is qualitatively analogous our neutral Children of Men scenarios (adults have low mortality and do not reproduce).

In the unstructured version of this model, species abundances will simply match their carrying capacities. In order to compare the effect of structure with our neutral results, we draw the carrying capacities from a logseries distribution, so that the SAD in the unstructured version of the model is logseries-distributed.

\begin{figure}[H]
\caption{Niche model results. \textbf{Top}: In the unstructured model (left column), abundances are log-series distributed, mirroring the distribution of carrying capacities. A log-series (blue curve) fits the SAD well in all scenarios regardless of stage structure. \textbf{Middle}: Unlike neutral dynamics, the progeny does not follow a power law. \textbf{Bottom}: Plotting $log(1+d_2N_2)$ against $log(1+N_1)$ reveals that adult and juvenile subpopulations are linearly related in all the structured scenarios. Red line: best fit; green line: 1-to-1 line.}\label{Niches}
\includegraphics[width=1\textwidth,angle=0]{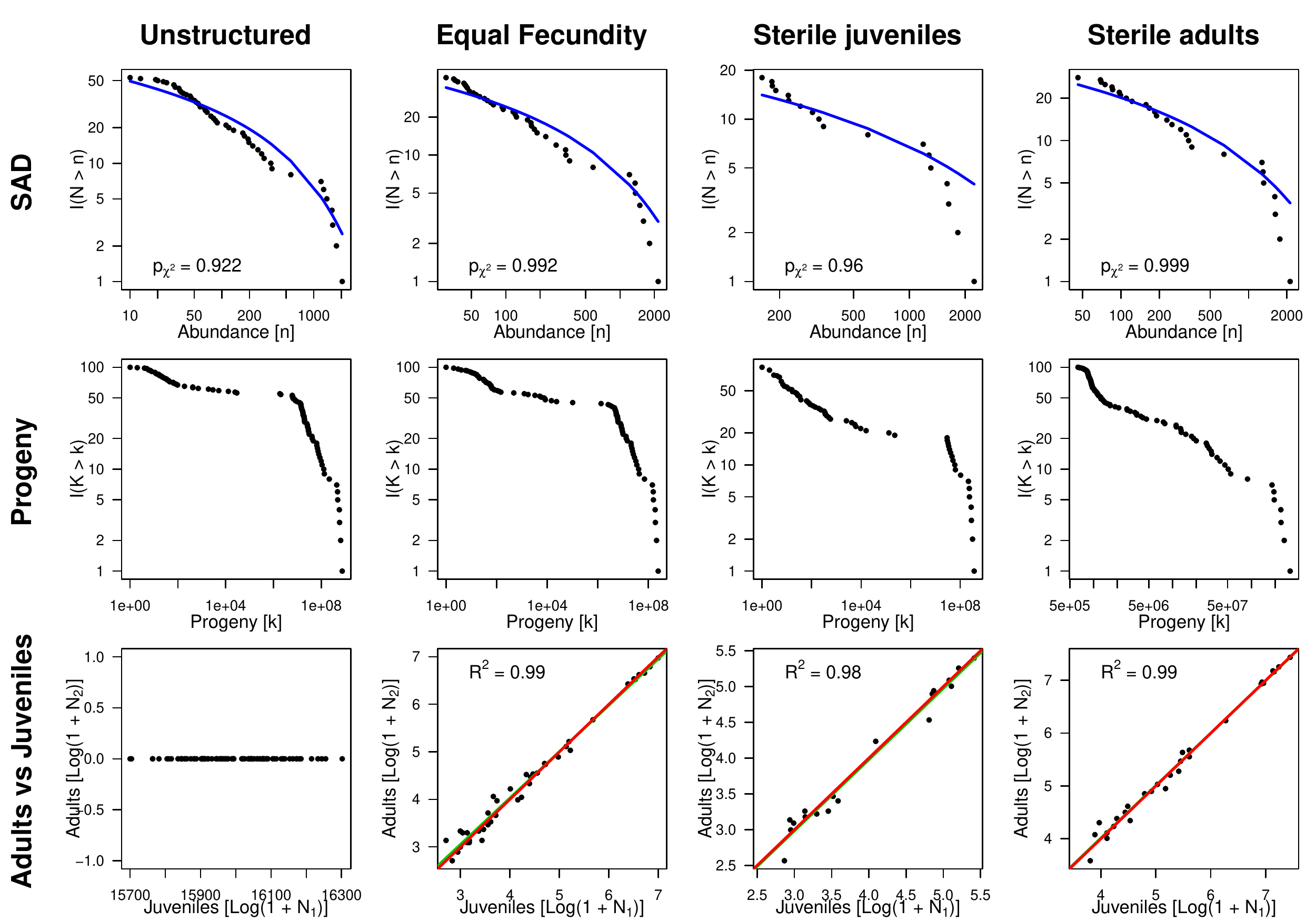}
\centering
\end{figure}

Results are shown in Figure \ref{Niches}. Like the neutral case, stage structure did not affect the SAD or the progeny distribution in the analogue to the Adult Vigor neutral scenario. However, neither distributions were affected even when adults had very low mortality and zero fecundity, in contrast with neutrality.

While the agreement between our models with and without species differences suggest that stage structure may not be a dominant driver of abundance and progeny distributions in general, the contrasting results in the case of low-mortality sterile adults may be due to the extreme niche structure in our model. Our results with the niche model presented here do not rule out the possibility that the SAD may still be informative of Children-of-Men-like stage structure in non-neutral dynamics with intermediate niche overlap between species.

\newpage
\section*{Appendix 3: Link between rank-abundance relationship and abundance distribution}
A species rank $r$ given its abundance $n_r$ is $r=S(N\geq n_r)$. This relates to the SAD as
\begin{equation*}
\langle S(N\geq n_r)\rangle=S\int_0^{n_r} P(N=n)\; \ud n
\end{equation*}
where $S$ is the total number of species. It follows that 
\begin{equation}
P(N=n_r)\propto\frac{\ud r}{\ud n_r}
\end{equation}
Therefore if the rank-abundance relationship is a power law $n_r\propto r^{-p}$ with exponent $p$, we conclude that the SAD must also be a power law $P(N=n)\propto n^{-b}$, with exponent $b=1+\frac{1}{p}$.


\newpage
\begin{figure}[H]
\caption{Time evolution of conceptual scenarios. \textbf{Top}: Total community size across simulation time (measured in number of events, ie births, deaths, aging, and speciation events throughout the community). \textbf{Middle}: Total juvenile population summed across species, $N_1=\sum_a N_{1a}$, maintains close connection with total adult population $N_2=\sum_a N_{2a}$, rarely deviating from the equilibrium condition $N_1(t)=d_2N_2(t)$ by more than 5\%. \textbf{Bottom}: Species richness across time.} \label{DiagPlotsMain}
\includegraphics[width=.8\textwidth,angle=0]{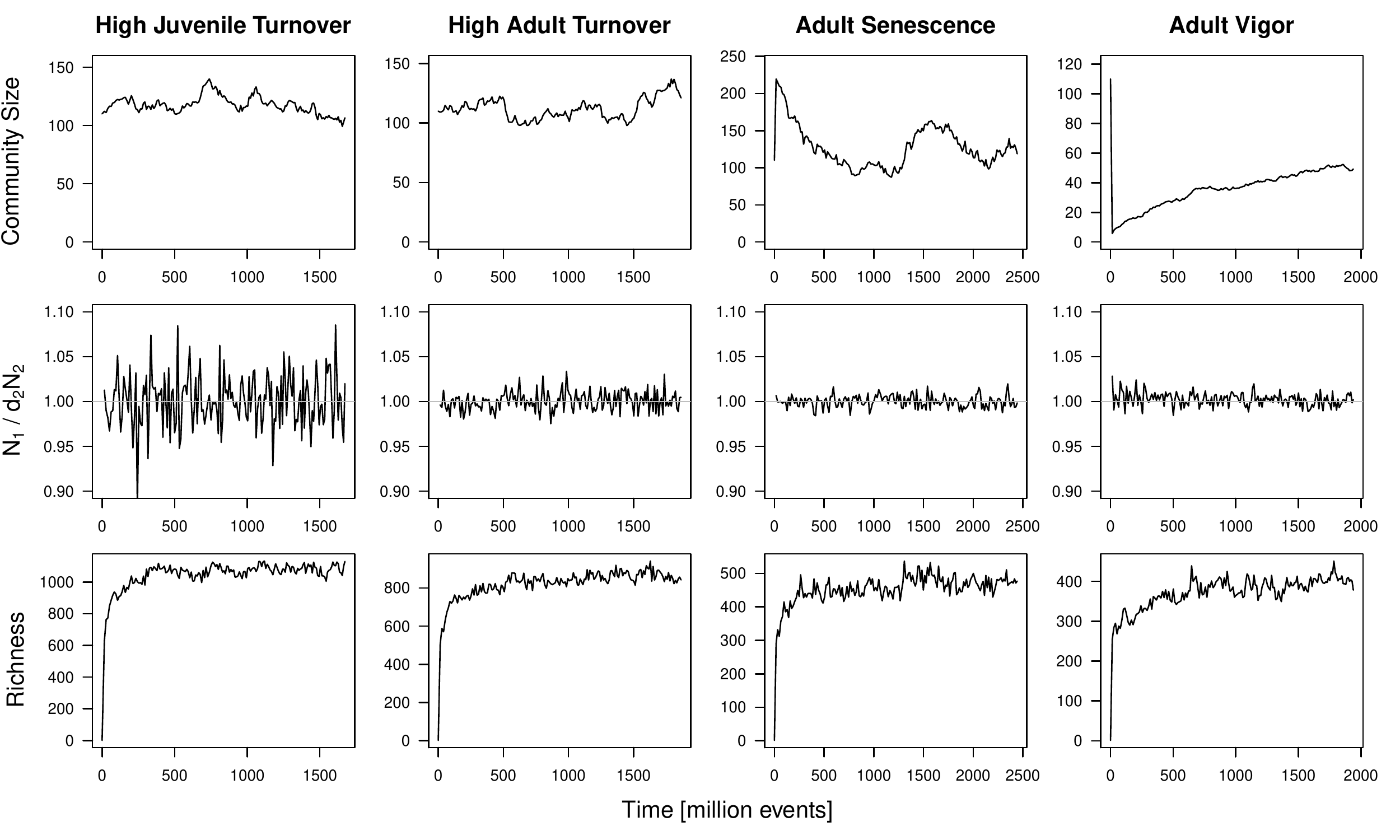}
\centering
\end{figure}

\newpage
\begin{figure}[H]
\caption{Children of Men scenarios. Plot layout similar to Figure \ref{DiagPlotsMain}. Notice much looser tie between juvenile and adult subpopulations as adult mortality becomes small (middle row). Wide fluctuations above 100\% departure from the equilibrium condition $N_1(t)=d_2N_2(t)$ become increasingly common for increasingly lower adult mortality. This occurs even though total community size and richness are in stationary equilibrium (top and bottom rows)} \label{DiagPlotsCOM}
\includegraphics[width=.8\textwidth,angle=0]{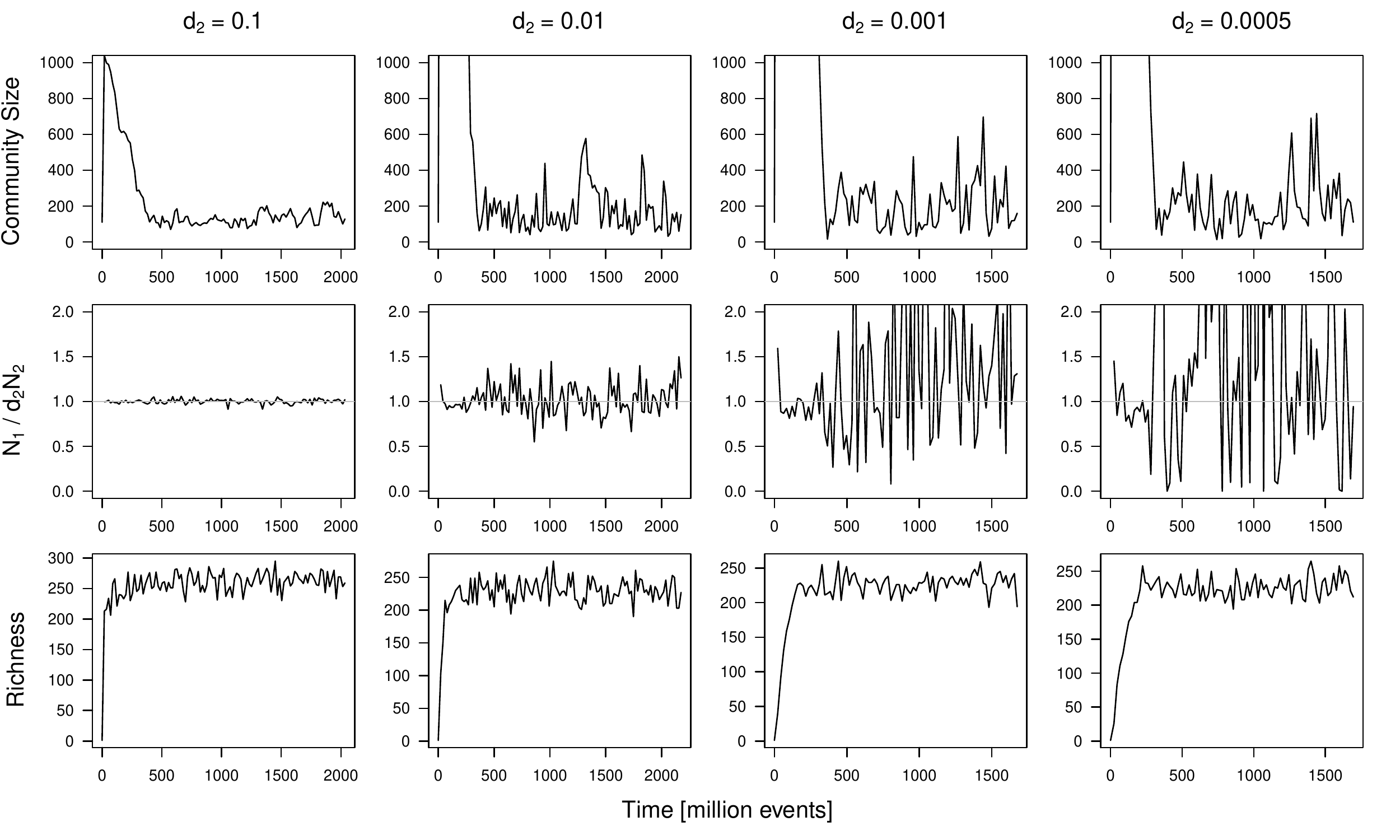}
\centering
\end{figure}

\newpage
\begin{figure}[H]
\caption{Eusocial insect scenarios. Plot layout similar to Figure \ref{DiagPlotsMain}. As in the Children of Men scenario, community size is highly variable compared with the first set of scenarios, and the subpopulations in the two life stages become disconnected as worker mortality gets low.} \label{DiagPlotsANT}
\includegraphics[width=.8\textwidth,angle=0]{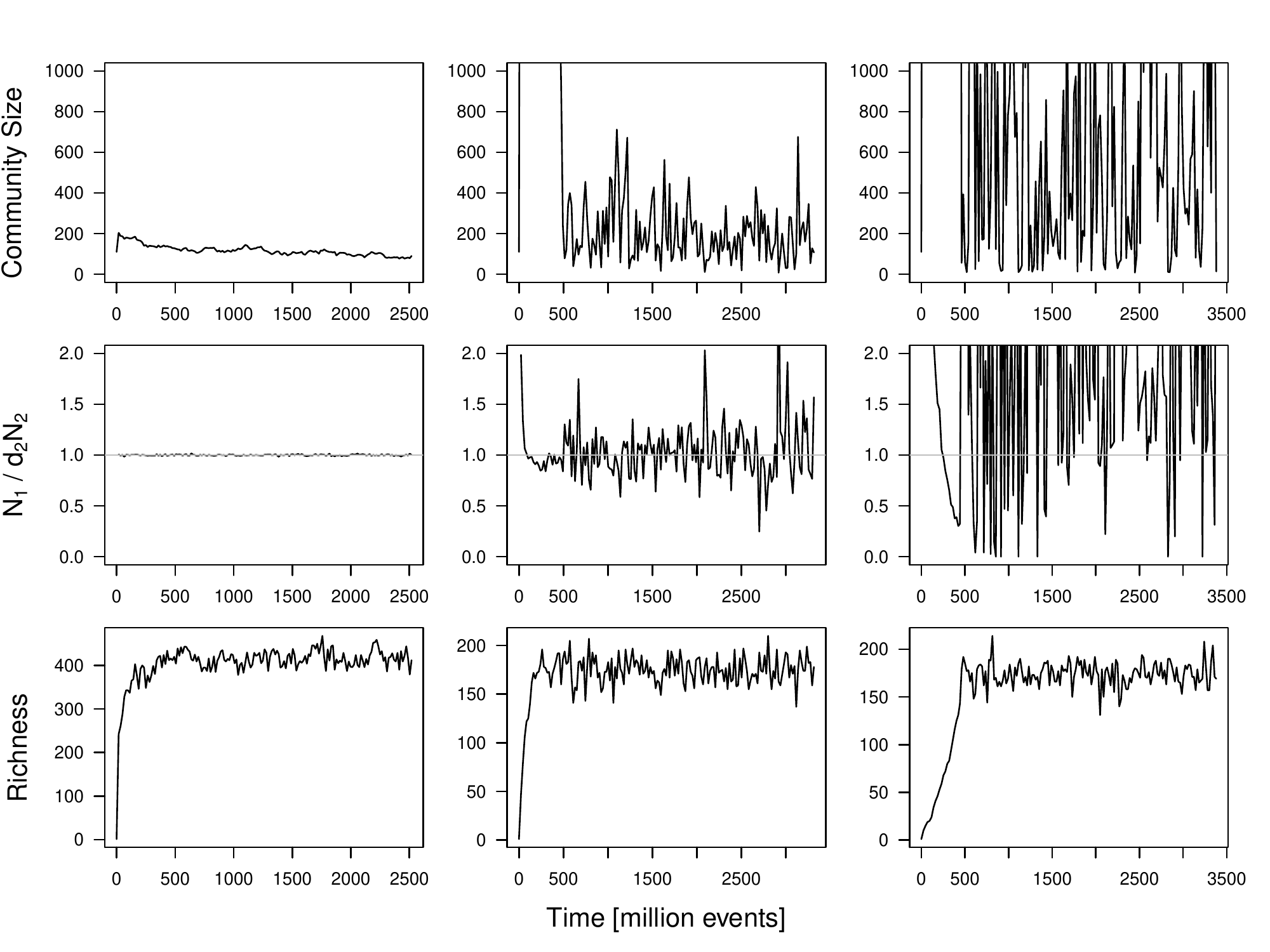}
\centering
\end{figure}

\newpage

\end{document}